# Tunable capacitance in all-inkjet-printed nanosheet heterostructures

Yang Wang[1], Mohammad Mehrali[1], Yi-Zhou Zhang[2], Melvin A. Timmerman[1], Bernard A. Boukamp[1], Peng-Yu Xu[1] and Johan E. ten Elshof[1*]

[1]University of Twente, MESA+ Institute for Nanotechnology, P. O. Box 217 7500AE Enschede, the Netherlands

[2]School of Chemistry and Materials Science, Nanjing University of Information Science and Technology, Nanjing 210044, China.

*Corresponding author. E-mail: j.e.tenelshof@utwente.nl

## Abstract

Heterostructures constructed from two-dimensional building blocks have shown promise for field-effect transistors, memory devices, photosensors and other electronic applications[1,2]. 2D nanosheet crystals can be constructed into multilayer heterostructures using layer-by-layer methods[3], but that method cannot be used to fabricate large-scale and thick heterostructures, due to the time-consuming nature and low efficiency of the process. An alternative approach to deposit different two-dimensional materials is by inkjet printing[4-7]. Here we show the fabrication of a nanosheet supercapacitor by inkjet printing $Ti_3C_2T_x$ MXene nanosheets as electrodes, and graphene oxide nanosheets as solid-state electrolyte. The free water molecules trapped between graphene oxide sheets facilitate proton movement through the layered solid electrolyte[8]. The as-made supercapacitor shows high areal capacitance, good cycling stability and high areal energy and power densities comparable with existing printed supercapacitors. Moreover, the specific capacitance can be increased further by addition of liquid electrolytes.

# Main

2D transition metal carbides or nitrides (MXenes) with general formula $M_{n+1}X_nT_x$ ($n = 1, 2, 3$), where M is an early transition metal, X is a carbon and/or nitrogen and $T_x$ are surface terminal groups like -F, -O, or -OH, have been attracting tremendous attention recently due to their outstanding chemical and physical properties[9,10]. MXenes with atomic thickness and high electrical conductivity have been widely studied for application in hydrogel sensors[11], solar cells[12] and supercapacitors (SCs)[13,14]. MXenes have been combined with other 2D materials into multi-material structures with tunable properties and functionalities, showing promise for energy storage applications[15]. On the other hand, hydrated graphene oxide (GO) nanosheets are electrically insulating but exhibit high ionic conductivity, suggesting their potential as solid-state electrolyte and separator[8].

MXene and GO are normally dispersed in a solvent like water, but traditional mechanical exfoliation and drying transfer techniques cannot be applied to fabricate MXene based heterostructures on large scale. Solution processing methods such as spray coating and vacuum filtration have been attempted, but these offer poor control over interface and surface roughness, resulting in poor device performance[15]. Inkjet printing, a simple, low-cost and versatile technique, provides an alternative route to the fabrication of large-scale vertical heterostructures with controlled thickness, interface and roughness[7]. Recently, various heterostructure devices based on printed 2D materials such as field-effect transistors[6], capacitors[16], photosensors and memory devices[7] have been demonstrated. However, realizing well controlled and sharp interfaces still presents a significant challenge for printed heterostructures. Full control over the heterostructure interface is key to achieving high performance, which includes avoiding redispersion of nanosheets from the interface upon deposition of a subsequent layer. The preparation of non-toxic, stable and printable 2D inks is another critical issue for inkjet printing. Owing to the wide range of physical properties present in 2D materials,[17] we demonstrate here that a combination of 2D materials can be used to realize an all-solid-state supercapacitor, without any liquid or gel electrolyte present in the system. In this work, we used a water-based additive-free MXene ink to inkjet print electrodes and current collectors on polyimide substrates, and a water-based GO ink to inkjet-print the solid-state electrolyte. Both sandwiched supercapacitors (SSCs) and micro-supercapacitor devices (MSCs) were printed on flexible polyimide substrates (Fig. 1). The SSCs achieved specific areal capacitances ($C_A$) up to 9.8 mF cm$^{-2}$ at a current density of 40 µA cm$^{-2}$. Addition of aqueous electrolytes led to enhancement of $C_A$, due to the improved ionic conductivity of the electrolyte resulting from the presence of additional ions and a liquid phase.

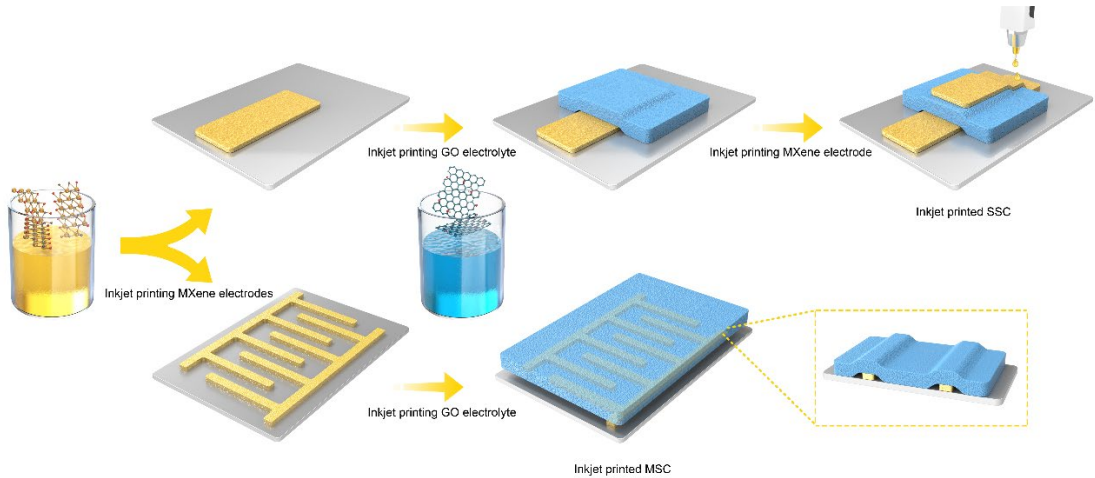

**Fig. 1 Schematic illustration of all-inkjet-printing-based heterostructure SSC (top) and MSC (bottom) supercapacitors.** Water-based additive-free MXene ink was first inkjet printed into thin films and interdigitated configurations as electrodes, followed by inkjet printing a water-based GO ink on top of interdigitated MXene electrodes to form an all-solid-state MSC (bottom). A second MXene electrode was inkjet printed on top of the solid-state GO electrolyte to complete the fabrication process of an all-solid-state SSC (above).

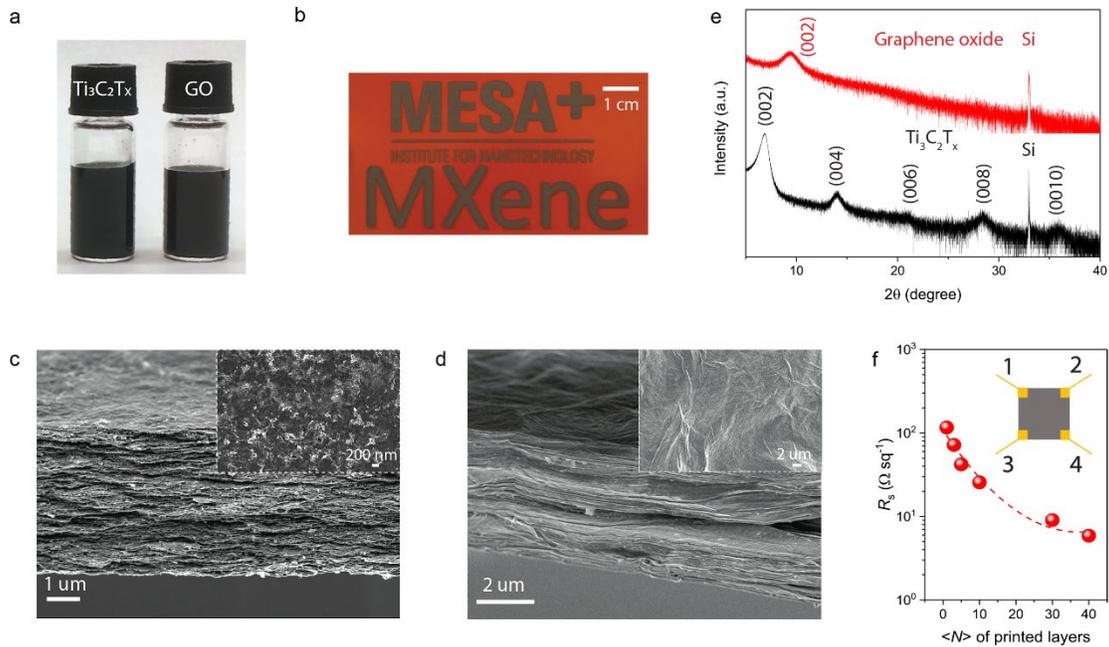

**Fig. 2 Characterization of MXene and GO nanosheets.** (**a**) Optical image of water-based MXene and GO inks. (**b**) Inkjet printed "MESA+ INSTITUTE FOR NANOTECHNOLOGY" logo and word "MXene" using MXene ink on polyimide substrate with multiple printing layers. Cross-sectional SEM image of (**c**) printed MXene and (**d**) GO films on Si/SiO$_2$ substrates. (**e**) XRD pattern of printed MXene and GO films on silicon substrates. (**f**) Sheet resistance $R_s$ as function of the number of printed MXene layers (5 by 5 cm$^2$) on Si/SiO$_2$. The inset shows a schematic representation of the sheet resistivity as measured by the Van der Pauw method. The ink concentration was around 4.5 mg mL$^{-1}$.

Additive-free water-based MXene and GO inks were successfully prepared as shown in Fig. 2a. Due to its high GO concentration, the GO ink has a dark brown color. The thickness of MXene and GO nanosheets were determined by atomic force microscopy (AFM) to be around 1.5 nm and 1 nm, respectively, indicating a unilamellar structure for both types (Supplementary Fig. S1a,b). The lateral sizes of MXene and GO nanosheets estimated from AFM images were about 0.76 μm for MXene nanosheets and 0.78 μm for GO nanosheets (Supplementary Fig. S1c). To evaluate the ink printability, the inverse Ohnesorge number $Z$, which is defined as $Z = (\alpha\rho\gamma)^{1/2}/\eta$, was employed. Here, $\alpha$ is the nozzle diameter, $\rho$ is the density, $\gamma$ is the surface tension, and $\eta$ is the viscosity of the fluid. The surface tension and viscosity of MXene ink were 80.3 mN m$^{-1}$ and 1.4 mPa s, respectively. The nozzle diameter of 21.5 μm and the value of surface tension, viscosity result in a $Z$ value of about 30 for MXene ink. The surface tension and viscosity of GO inks were 129.4 mN m$^{-1}$ and 20.2 mPa s, respectively, leading to a $Z$ value of about 3. Stroboscopic images of ink droplet formation versus time illustrated the quality of the inks (Supplementary Fig. S4). No satellite droplets were observed for MXene and GO inks. Both inks showed good wetting on silicon substrates, as confirmed by AFM mappings, and the cross-sectional profile of the AFM images further confirmed the uniform deposition of both MXene and GO inks (Supplementary Fig. S5). The wrinkles in printed GO droplets were caused by interactions between adjacent GO sheets[18]. Fig. 2b shows two examples of printed patterns obtained by MXene ink on flexible polyimide substrate, demonstrating flexibility in pattern design and large area coating with multiple printing passes. The scanning electron microscopy (SEM) images of printed MXene (Fig. 2c) and GO (Fig. 2d) nanosheet films on Si/SiO$_2$ substrates show uniformity and continuity over large surface areas. It is worth noting that the sheets in both printed films showed a high degree of horizontal orientation and a layer-by-layer structure, which will facilitate the transport of electrolyte ions in in-plane structured devices such as MSCs. As shown in Fig. 2e, the XRD pattern of a printed MXene film shows strong ordering in the $c$ direction with a (002) peak at 6.8°, thus confirming the horizontal orientation of nanosheets in printed films. The smaller angle than in dry MXene films, where the same peak is at 8.9° (Supplementary Fig. S2a), indicates wider spacing between the layers in the printed film and intercalation of spatially confined H$_2$O molecules[19]. The XRD pattern of a printed GO film shows a peak at 2θ = 9.6°, which corresponds with a $d$ spacing of 0.92 nm, suggesting that electrolyte ion transport is predominant in horizontal rather than in vertical direction. The sheet resistance $R_s$ of printed MXene films could be tuned by the number of printed layers. As shown in Fig. 2f, the $R_s$ of MXene films on Si/SiO$_2$ substrates decreased rapidly from 116.7 Ω sq$^{-1}$ (printed layers $<N> = 1$) to around 5.9 Ω sq$^{-1}$ ($<N> = 40$) with an ink concentration of around 4.5 mg ml$^{-1}$.

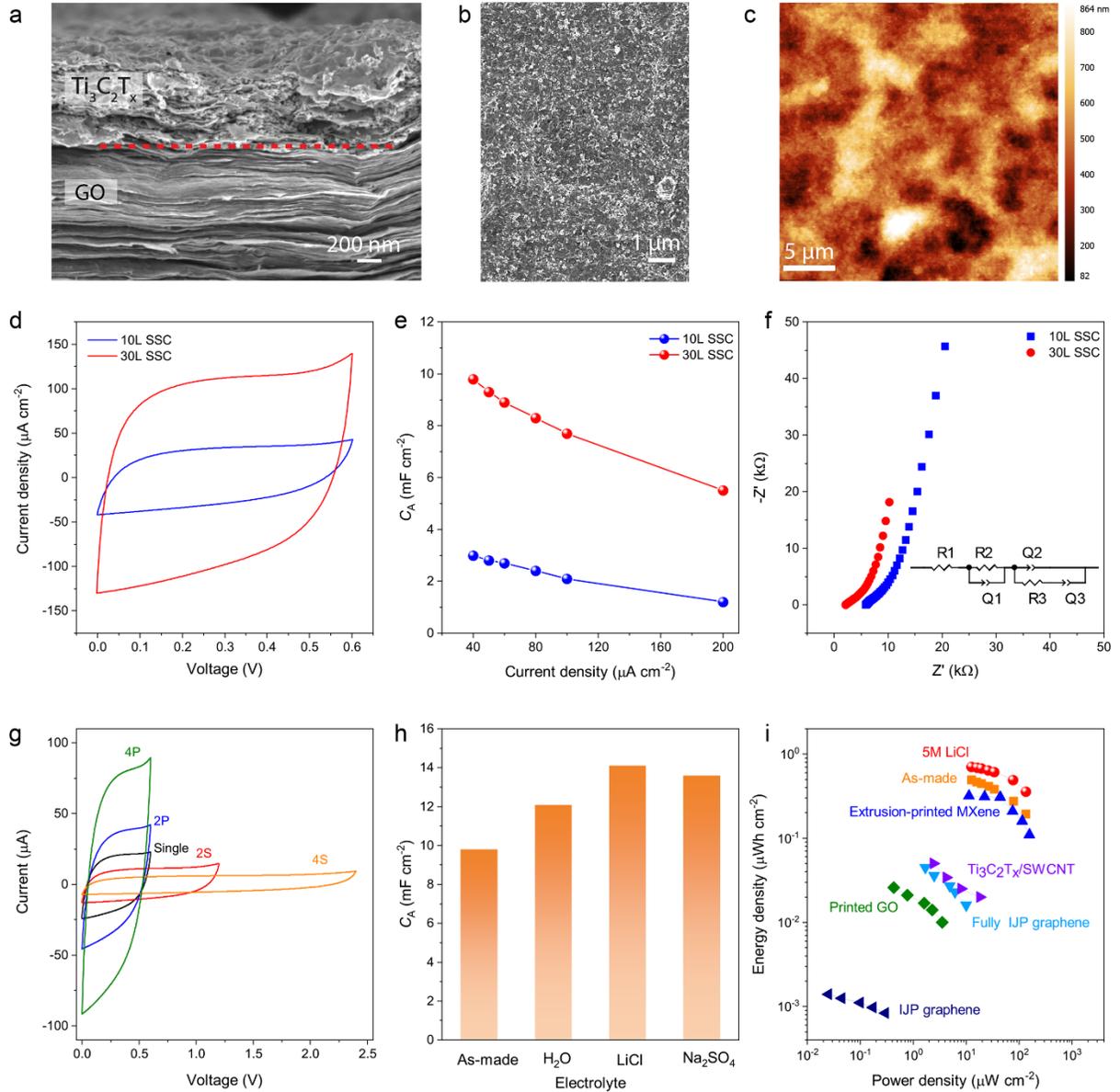

**Fig. 3 Electrochemical performance of all-inkjet-printed SSCs.** (**a**) Cross-sectional SEM image of inkjet-printed SSC. The dashed line roughly indicates the boundary between MXene (above) and GO (below) phases as a guide to the eye. (**b**) Top view SEM of top MXene electrode. (**c**) AFM topography of 30×30 μm$^2$ scan area in device. (**d**) CV of as-made 10L SSC and 30L SSC at 10 mV s$^{-1}$. (**e**) $C_A$ of as-made 10L SSC and 30L SSC at different current

densities. (**f**) Nyquist plots of as-made 10L SSC and 30L SSC. The inset shows the equivalent circuit to which the experimental data were fitted. (**g**) CV diagram of 1, 2 and 4 supercapacitors connected in series and in parallel. (**h**) $C_A$ of 30L SSC with different electrolytes (DI water, 5 M LiCl and 0.5 M $Na_2SO_4$ in water, respectively) on top of the devices. **i**, Ragone plot of all-inkjet-printed SSC with other different systems.

The solid-state symmetrical SSCs consist of a printed solid-state GO electrolyte sandwiched between two printed MXene electrodes (Fig. 1, Supplementary Fig. S6a,c). A clear boundary between electrolyte and electrodes can be identified, demonstrating well-defined spatial separation between the phases (Supplementary Fig. S6b). To avoid remixing of nanosheets at the printed interfaces, printed nanosheet layers were solidified by heating at 50 °C for 1h before printing another material on top. Cross-sectional SEM analysis indicated an intimate and stable contact between the MXene electrodes and the GO electrolyte (Fig. 3a, Supplementary Fig. S6d-f). The top-view SEM image of a MXene electrode shows continuous features of printed electrodes without cracks or pin holes, illustrating the high quality of printed films (Fig. 3b). Due to the high resolution brought by inkjet printing, all printed 30L SSCs exhibited a low root mean square roughness (RMS) of ≈ 130 ± 25 nm at a device thickness of around 4 μm (Fig. 3c). Element mappings from energy-dispersive x-ray spectroscopy (EDS) further confirm the stable and sharp interface between MXene electrodes and GO electrolyte (Supplementary Fig. S7). To investigate the electrochemical performances of all-inkjet-printed SSCs, MXene electrodes with thicknesses of 10 and 30 printed layers (from here on referred to as 10L SSC and 30L SSC, respectively) were fabricated. As shown in Fig. 3d, 30L SSC shows a higher specific capacitance than 10L SSC from cyclic voltammetry (CV) at scan rate of 10 mV s$^{-1}$. The quasi-rectangular CV curves show the pseudo-capacitive behavior of the devices. More specifically, 30L SSC still exhibits a quasi-rectangular shape even at high scan rate (Supplementary Fig. S8a,c).

Galvanostatic charge/discharge (GCD) data of both devices at current densities ranging from 40-200 µA cm$^{-2}$ (Supplementary Fig. S8b,d) are shown in Fig. 3e. The 30L SSC exhibits $C_A$ as high as 9.8 mF cm$^{-2}$ at current density of 40 µA cm$^{-2}$, while 10L SSC exhibits $C_A$ of 3 mF cm$^{-2}$ at same current density, indicating that $C_A$ is roughly proportional with the number of printed MXene layers.

Both types of devices were characterized without liquid electrolyte. The mobile electrolyte ions needed for charging/discharging these devices are therefore thought to arise from the hydrolysis of functional oxygen-bearing groups on the solid GO electrolyte[8]. Free water molecules present between GO sheets may facilitate proton transport via the Grotthuss mechanism[20] or by diffusion of hydronium ions within the interlayer spaces[8]. Intimate contact between the GO electrolyte and MXene electrodes will facilitate proton transfer between different SSC components. Proton movement inside the MXene electrodes probably proceeds via confined water molecules that are trapped between MXene sheets.

Electrochemical impedance spectroscopy (EIS) was conducted on both devices in the frequency range from 10 mHz to 10 kHz. The experimental data were fitted to the equivalent circuit shown in the inset of Fig. 3f. It consists of a constant phase element (CPE) $Q_2$ that represents the surface capacitance of the device, in parallel with a CPE $Q_3$ that represents the slower diffusion-controlled volume capacitance. The electrolyte is represented by the $R_2(Q_1)$ sub-circuit, and the electrode resistance by $R_1$. Sample 10L SSC has a larger $R_1$ than 30L SSC, showing that thicker electrodes exhibit a lower resistance (Table S1; 5.9 kΩ versus 2.1 kΩ). The impedance in the low frequency range suggests mixed surface absorption and diffusional control of the devices, which are the double layer capacitance ($Q_2$) and charge transfer diffusion impedance $Q_3$. The 30L SSC device has a smaller charge transfer resistance ($R_3$) in the lower frequency range than 10L SSC

(1.9 kΩ versus 12.9 kΩ).

To demonstrate the potential for practical applications at high voltages, the as-made 30L SSCs were connected in series and in parallel configurations. As shown in Fig. 3g, the voltage window reached 1.2 V and 2.4 V with two and four devices connected in series, respectively. The current was increased by a factor of ~ 2 and 4 with two and four devices connected in parallel, respectively. The electrochemical performance of SSCs could be enhanced by addition of liquid electrolytes. As shown in Fig. 3h, excess deionized (DI) water resulted in a higher $C_A$ of 12.1 mF cm$^{-2}$ than in the as-made device, probably due to enhanced ion transport in liquid media. Aqueous electrolytes such as 5 M LiCl and 0.5 M $Na_2SO_4$ introduce additional electrolyte ions that enhance ionic transport at the same time, increasing the $C_A$ further to 13.6-14.1 mF cm$^{-2}$ (Supplementary Fig. S9). The as-made 30L SSC exhibited an areal energy density ($E_A$) of 0.49 µWh cm$^{-2}$ at a power density ($P_A$) of 12.55 µW cm$^{-2}$, while $E_A$ increased to 0.71 µWh cm$^{-2}$ at $P_A$ of 12.48 µW cm$^{-2}$ by adding a drop of 5M LiCl electrolyte into the as-made device. Apparently, the presence of liquid electrolyte has a positive effect on the energy density, however the precise mechanism remains unclear. The $E_A$ of as-made 30L SSC is higher than the recently reported extrusion-printed MXene MSC with $H_2SO_4$-poly(vinyl alcohol, PVA) gel electrolyte ($E_A$ of 0.32 µWh cm$^{-2}$ at $P_A$ of 11.40 µW cm$^{-2}$)[21], fully printed MSC with GO electrolyte[22], all printed MSC with graphene electrodes and printed polyelectrolyte[23], printed graphene MSC[24], and MXene/single-walled carbon nanotube supercapacitor[25]. The 30L SSC exhibits good cycling stability with a capacitance retention of ~100% after 10000 cycles (Supplementary Fig. S10a). Moreover, the all-printed SSC shows high mechanical stability with a bending radius of about 1 cm, as shown in Supplementary Fig. S10b.

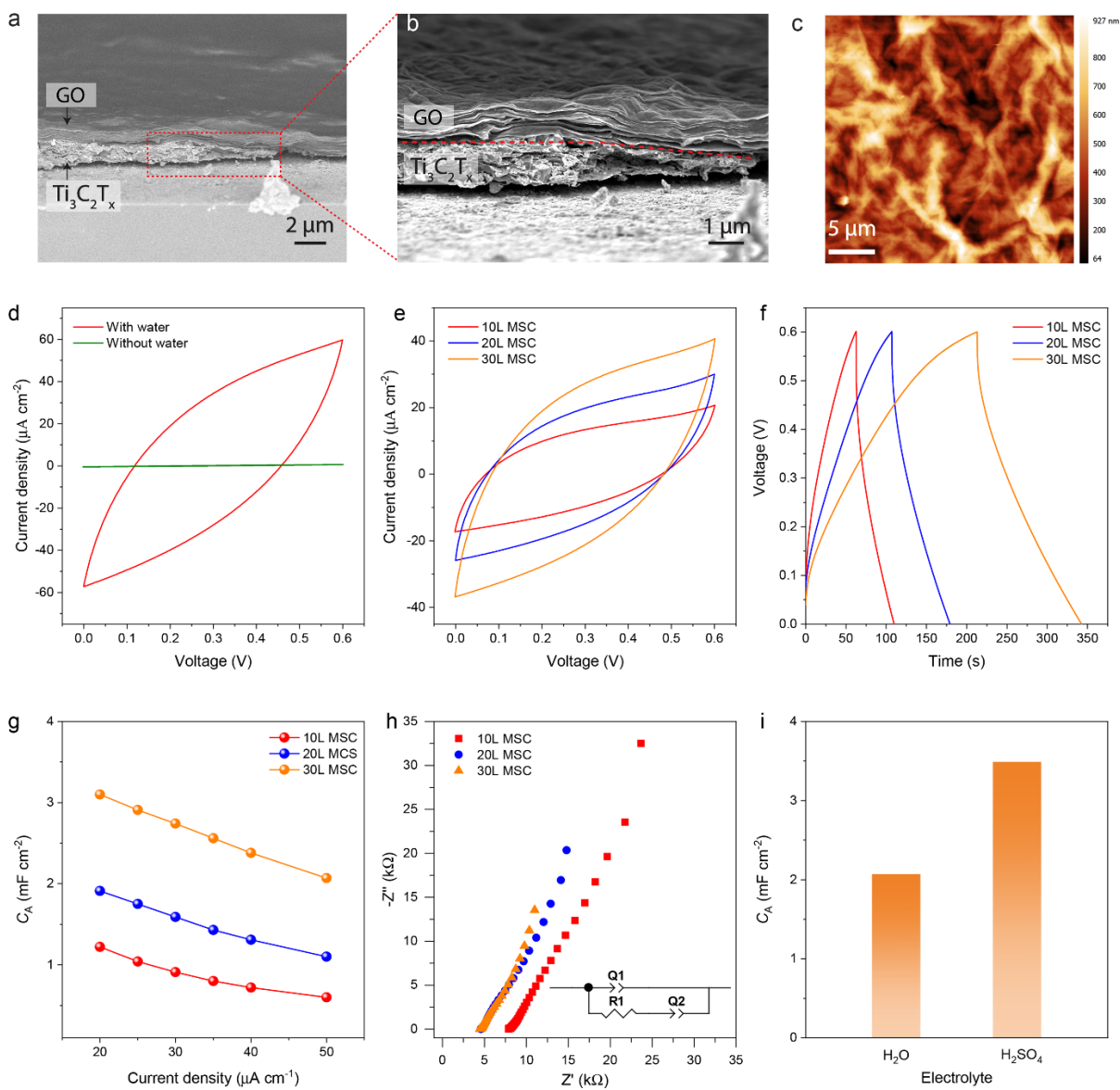

**Fig. 4 Electrochemical performance of all-inkjet-printed MSCs.** (**a, b**) Cross-sectional SEM images of all-inkjet-printed MSC. The dashed line roughly divides MXene electrode (below) and GO electrolyte (above). (**c**) AFM topography of top GO electrolyte in (**b**). (**d**) CV of 30L MSC with and without added water on top of the device at 20 mV s$^{-1}$. (**e**) CV of printed 10L/20L/30L MSCs at 10 mV s$^{-1}$. (**f**) GCD of printed 10L and 30L MSCs at 15 µA cm$^{-2}$. (**g**) $C_A$ of printed 10L/20L/30L MSCs with different current densities. (**h**) Nyquist plots of printed 10L/20L/30L MSCs. The inset is the fitted equivalent circuit for the low frequency range of the impedance response. (**i**) $C_A$ of 30L MSC with excess H$_2$O and 0.5 M H$_2$SO$_4$ at current density of 50 µA cm$^{-2}$.

MSCs were fabricated by printing MXene nanosheets with interdigitated structure as electrodes on polyimide substrate, followed by a printed layer of GO nanosheets on top of/over the MXene electrodes to serve as solid-state electrolyte (Fig. 1, bottom). MXene electrodes with 10, 20 and 30 printed layers (refers to as 10L MSC, 20L MSC and 30L MSC, respectively) were printed (Supplementary Fig. S11). The printed MXene electrodes show sharp features at the edges, indicating the stability of the MXene ink and the high reliability of the inkjet printing process, and the sharp edges were still retained after printing GO nanosheets on top. EDS elemental mapping confirms the presence of a well-defined and stable top interface between MXene electrode and GO electrolyte (Supplementary Fig. S12). The all-inkjet-printed 30L MSCs exhibited a low RMS of $\approx 157 \pm 20$ nm at a device thickness of around 2.3 µm (Fig. 4c). However, the contact at the cross-section between MXene electrode and GO electrolyte is poor (Fig. 4a,b), which is likely the cause of low current response in the CV measurement (Fig. 4d). Most likely, only GO sheets near the heterostructure interface contribute to the capacitance, because the protons are generated via hydrolysis of functional groups on GO. Electrochemical measurements were performed on as-made all-solid-state MSCs and on MSCs to which excess aqueous electrolyte had been added. The electrochemical performance of 30L MSC improved considerably upon addition of water, due to enhanced proton mobility (Fig. 4d). The CV curves of MSCs with varying electrode thicknesses demonstrate that thicker electrodes with more active surface sites show higher $C_A$ (Fig. 4e, Supplementary Fig. S13a-c). GCD further confirms that 30L MSC exhibits a higher capacitance than the other two devices (Fig. 4f, Supplementary Fig. S13d-f). The $C_A$ of 30L MSC reached to 3.1 mF cm$^{-2}$, while 10 and 20 layer devices reached 1.2 and 1.9 mF cm$^{-2}$ at a current density of 20 µA cm$^{-2}$, respectively (Fig. 4g). EIS suggests that the charge transfer resistance ($R_1$) of the 30 layers thick MXene electrode device is lower than the other two devices (4.1 kΩ, 11.9 kΩ and 12.0 kΩ for10L SSC, 20LSSC and 30L SSC, respectively; Fig. 4h, Table S2). Similar to the SSC, the equivalent circuits in the low frequency range suggests mixed surface absorption and diffusional control, i.e. the double layer capacitance ($Q_1$) and the charge transfer diffusion impedance ($Q_2$). Addition of a 0.5 M $H_2SO_4$ electrolyte solution onto 30L MSC resulted in a a higher capacitance than in devices with excess water (Fig. 4i, Supplementary Fig. S14). The $H_2SO_4$ electrolyte provides additional protons that enhance the ionic conductivity, leading to lower series resistances.

In conclusion, we demonstrated all-inkjet-printed solid-state supercapacitors based on 2D MXene/GO/MXene. Due to the high ionic conductivity of GO films, the printed SSC without liquid electrolyte showed an $E_A$ of 0.49 μWh cm$^{-2}$ at a $P_A$ of 12.55 μW cm$^{-2}$. The $C_A$ can be increased further by adding liquid electrolyte. Further optimization on materials, printed electrodes thickness and electrodes configuration will enhance device performance further. Printed supercapacitors show high potential for use in small power source units for flexible electronics.

**Acknowledgements**

Y.W. acknowledges the financial support of the China Scholarships Council program (CSC, No. 201608340058). Y.Z.Z. acknowledges the financial support from the National Natural Science Foundation of China (21805136) and the Natural Science Foundation of Jiangsu Province (BK20170999). Y.W. thanks Dr. B. Chen for the help with electrical conductivity measurements and Dr. Y. Liu for the help with surface tension measurements. M. Smithers is acknowledged for performing the HR-SEM and EDS elemental mapping.

**Author Contributions**

Y.W. and J.E.t.E. conceived the research. Y.W. prepared the inks, made devices, did materials characterization and all electrochemical measurements with the assistance from Y.Z.Z. Y.W. synthesized the GO nanosheets with the help of M.A.T. M.M. synthesized MXene nanosheets and performed XRD of MAX and MXene. Y.W. and P.Y.X. performed Raman measurements and analyzed the data. B.A.B. conducted EIS analysis. Y.W. and J.E.E. wrote the manuscript. All authors discussed the results and commented on the manuscript.

**Competing interests statement**

The authors declare no competing interests.

**Additional information**

**Supplementary information** is available in the online version of the paper.

**Correspondence and requests for materials** should be addressed to J.E.E.

# Supplementary Information

## Tunable capacitance in all-inkjet-printed nanosheet heterostructures

Yang Wang[1], Mohammad Mehrali[1], Yi-Zhou Zhang[2], Melvin A. Timmerman[1], Bernard A. Boukamp[1], Peng-Yu Xu[1] and Johan E. ten Elshof[1*]

[1]University of Twente, MESA+ Institute for Nanotechnology, P. O. Box 217 7500AE Enschede, the Netherlands

[2] School of Chemistry and Materials Science, Nanjing University of Information Science and Technology, Nanjing 210044, China.

*Corresponding author. E-mail: j.e.tenelshof@utwente.nl

**Supplementary Materials:**

Experimental section

Supplementary Figures 1-14

Supplementary Table 1-2

## Experimental section

Titanium carbide ($Ti_3C_2T_x$) MXene was synthesized following a mild etching method as outlined elsewhere[1]. Typically, the etchant solution was prepared by dissolving 3.2 g of lithium fluoride (LiF, Sigm-Aldrich, −300 mesh powder, 98.5%) in 40 mL of 9M HCl (Sigm-Aldrich, 37% solution in water). Subsequently, 2 g of sieved $Ti_3AlC_2$ powder (400 mesh) was slowly added to the etchant solution over the course of 10 min and the reaction temperature was kept at 35 °C. After reaction for 24 h, the resultant was washed with deionized water repeatedly and delaminated manually by hand shaking agitation to obtain $Ti_3C_2T_x$ MXene suspension. The prepared solution was stored in a nitrogen-sealed vial and used as the MXene ink.

**Preparation of graphene oxide (GO) nanosheets ink**

Graphite oxide was synthesized from natural graphite (Nord-Min 802, Chemical Schmits Solutions) by a modified Hummers method[2]. Graphite (2 g) was added to 50 mL concentrated sulfuric acid (Fluka) in a 1000 mL flask under stirring in an ice bath for 2 h. Then 7 g potassium permanganate (Merck) was slowly added to the suspension under vigorous stirring to keep the temperature of the mixture under 10 °C. The mixture was transferred to a 35 °C oil bath under stirring for 20 h, yielding a thick paste. After the mixture had cooled down to room temperature, 100 mL DI water was slowly added with vigorous stirring for 2 h while keeping the flask in an ice bath. An additional 500 mL DI water was added, followed by addition of 15 mL $H_2O_2$ (30 wt%, Aldrich) until no further bubbles came out. The mixture was washed by 1 : 10 HCl (37%, Acros organics) solution (250 mL) to remove metal ions, and subsequently with DI water to a pH around 6. The resulting solid was freeze dried. The freeze dried graphite oxide powder was dispersed in DI water by ultrasonication for 2 h to get a GO suspension. To increase the GO concentration, the GO suspension was centrifuged at 15000 g for 1 h. The collected sediment was re-dispersed in printing solvent containing 0.06 wt% Triton X-100 (Sigma-Aldrich) and 1:10 propylene glycol (Sigma-Aldrich) : water by mass.

**Inkjet printing**

All patterns and devices were inkjet printed by a Dimatix DMP-2800 inkjet printer (Fujifilm Dimatix), which was equipped with a 10 pL cartridge (DMC-11610). To perform AFM measurements on single droplets, the MXene and GO inks were inkjet printed on $Si/SiO_2$ with a drop spacing of 80 μm at 30 °C. For the electrical conductivity measurements, the MXene ink was printed at 30 °C on $Si/SiO_2$ as a thin film with size 5 mm by 5 mm and varying numbers of layers at a drop spacing of 20 μm.

To fabricate all-inkjet-printed solid-state sandwiched supercapacitors, the MXene ink was first printed at 30 °C as bottom electrode on polyimide substrate with a drop spacing of 20 μm followed by drying at 50 °C for 1 h. Then GO ink was printed on top of MXene electrode with a drop spacing of 20 μm at 30 °C followed by drying at 50 °C for 1 h. Finally, the MXene ink was printed at 30 °C on GO electrolyte as top electrode with a drop spacing of 20 μm. It is worth noting that the GO electrolyte area is larger than the MXene electrode to prevent short-circuiting. To make all-inkjet-printed solid-state micro-supercapacitors, the MXene ink was printed with an interdigitated configuration on polyimide substrate as interdigitated electrodes at 30 °C with different layers at a drop spacing of 20 μm using 2 nozzles, followed by drying at 50 °C for 1 h. Then, the GO ink was printed at 30 °C on top of the MXene electrodes at a drop spacing of 20 μm.

**Electrochemical characterization**

All electrochemical characterizations were conducted on an Autolab workstation (PGSTAT128 N). Both micro-supercapacitors and sandwiched supercapacitors were characterized in a two-electrode configuration. Electrochemical impedance spectroscopy was performed by applying an AC voltage of 10 mV amplitude in the frequency range from 0.01 Hz to 10 kHz.

**Materials characterization**

X-ray diffraction (XRD) analysis was done with a PANalytical X'Pert Pro with Cu Kα radiation (λ=0.15405 nm). Atomic Force Microscopy (AFM) (Veeco Dimension Icon) was conducted in standard tapping mode. The AFM data were analyzed by Gwyddion (version 2.47) software. X-ray photoelectron spectroscopy (XPS) was conducted using an Omicron Nanotechnology GmbH (Oxford Instruments) surface analysis system with a photon energy of 1486.7 eV (Al Kα X-ray source) with a scanning step size of 0.1 eV. The pass energy was set to 20 eV. The spectra were corrected using the binding energy of C 1$s$ of the carbon residual on nanosheets as a reference. Raman spectroscopy was performed on a Bruker Senterra Raman spectrometer using a 532 nm laser under ambient conditions. High resolution scanning electron microscopy (HRSEM; Zeiss MERLIN) was performed to acquire information of printed MXene and GO films. The surface tension of the inks was measured by a contact angle system OCA (Data Physics Corporation). The viscosity of the inks was determined by an Automated Microviscometer AMVn (Anton Paar GmbH).

The electrical conductivity of printed MXene films on Si/SiO$_2$ was measured in a Van der Pauw geometry by Physical Properties Measurement System (PPMS) at 300 K. Copper wires were bonded on four corners of printed MXene films by silver paste. $R_s$ was calculated from the following Equation (1):

$$R_S = \frac{\pi R}{\ln 2} \quad (1)$$

The specific areal capacitance ($C_A$) of film electrodes was calculated from the GCD curves by using Equation (2):

85     $C_A = [I/(dV/dt)]/A_{electrode}$      (2)

86     where $I$ is the discharge current, $dV/dt$ is the slope of discharge curve, and $A_{electrode}$ refers to the

87     geometrical surface area of the film electrode.

88     The specific areal capacitance ($C_{A,device}$) of the MSC devices were also calculated from the GCD

89     curves using equations (3):

90     $C_{A,\,device} = [I/(dV/dt)]/A_{device}$      (3)

91     Here $A_{device}$ refers to the total geometrical surface area of the device including the electrodes and

92     the gap between the electrodes.

93     The areal energy densities ($E_A$, μWh cm$^{-2}$) and power densities ($P_A$, μW cm$^{-2}$) were calculated

94     from equations (4) and (5)

95     $E_A = C_{A,\,device} V^2/(2 \times 3.6)$      (4)

96     $P_A = 3600 \times E_A/\Delta t$      (5)

97     Where $\Delta t$ refers to discharge time.

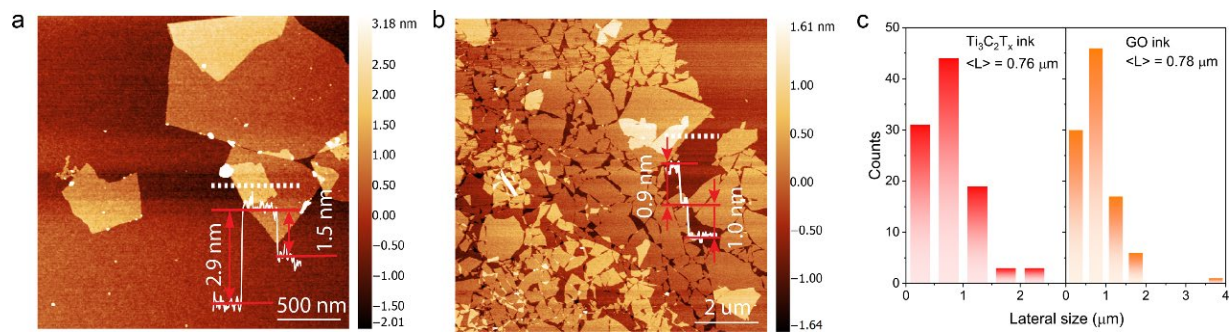

**Fig. S1 MXene and GO sheet thicknesses and lateral sizes.** AFM images of (a) $Ti_3C_2T_x$ and (b) GO nanosheets. (c) Lateral size distribution of $Ti_3C_2T_x$ and GO nanosheets estimated by measuring 100 flakes dropcast on $Si/SiO_2$.

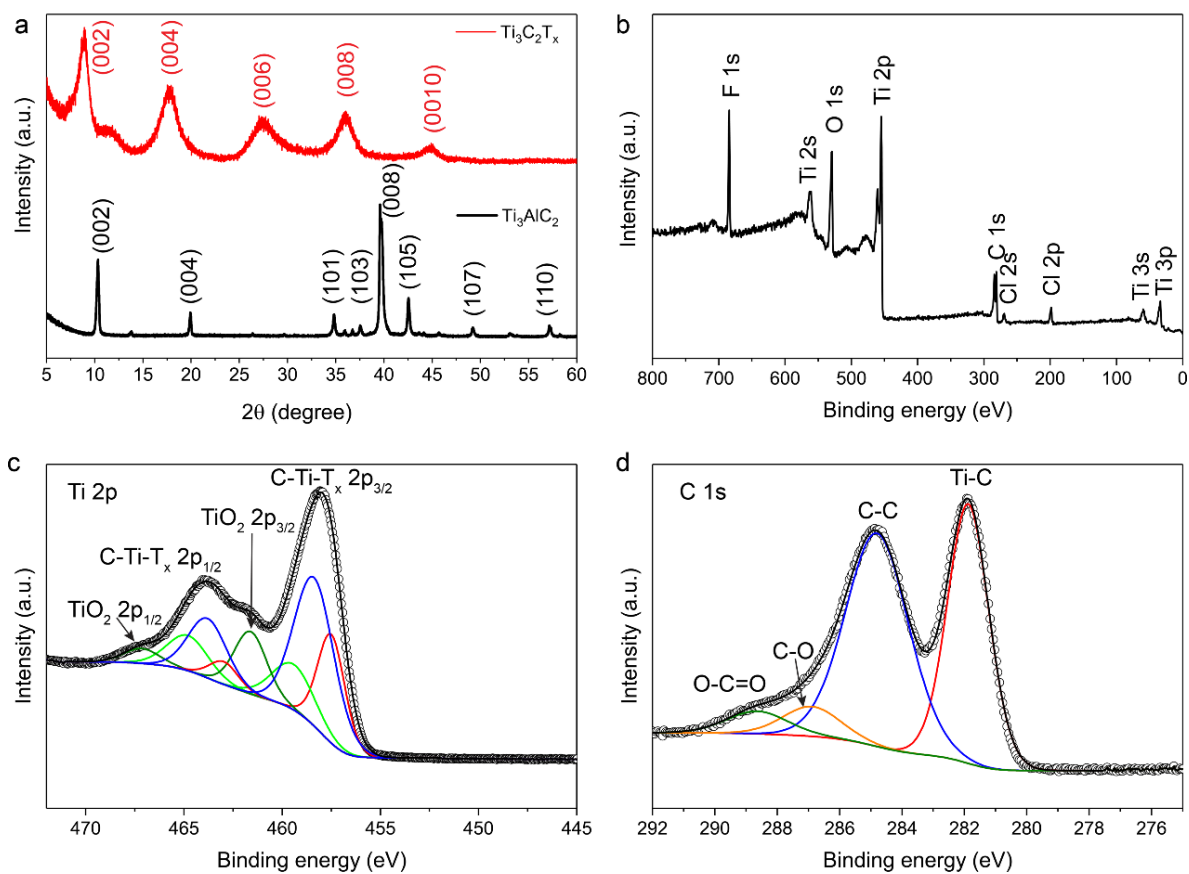

**Fig. S2 Characterization of Ti$_3$C$_2$T$_x$ nanosheets.** (a) XRD diagram of Ti$_3$AlC$_2$ powder and Ti$_3$C$_2$T$_x$ nanosheets; (b) XPS survey spectrum of Ti$_3$C$_2$T$_x$ nanosheets; High resolution XPS spectrum of (c) Ti 2$p$ and (d) C 1$s$ of Ti$_3$C$_2$T$_x$ nanosheets.

As shown in Fig. S2a, Ti$_3$C$_2$T$_x$ nanosheets show larger interlayer spacings than Ti$_3$AlC$_2$ due to the successful removal of Al, as can be concluded from the smaller diffraction angle of the (002) peak of the former. No Al peak was identified in the XPS survey spectrum, which further confirms the successful exfoliation of Ti$_3$C$_2$T$_x$ nanosheets (Fig. S2b). The high-resolution XPS spectrum of Ti 2$p$ could be fitted with four doublets (Ti 2$p_{3/2}$-Ti 2$p_{1/2}$). The Ti 2$p_{3/2}$ components located at 457.5, 458.4 and 459.4 eV correspond to Ti-C, Ti (II) and Ti (III), respectively (Fig. S2c). The peak at 461.6 eV belongs to Ti 2$p_{3/2}$ of TiO$_2$. The C 1$s$ is fitted by four components

114    corresponding to Ti-C at a binding energy (BE) of 281.9 eV, C-C at 284.8 eV, C-O at 286.9 eV

115    and O-C=O at 288.6 eV (Fig. S2d). The Ti-C peak is from MXene, while the other three peaks

116    are from MXene surface termination group originated from chemical etching. The high

117    resolution XPS spectra of Ti $2p$ and C $1s$ reveal that the MXene surfaces are terminated by F, O

118    and OH functional groups.

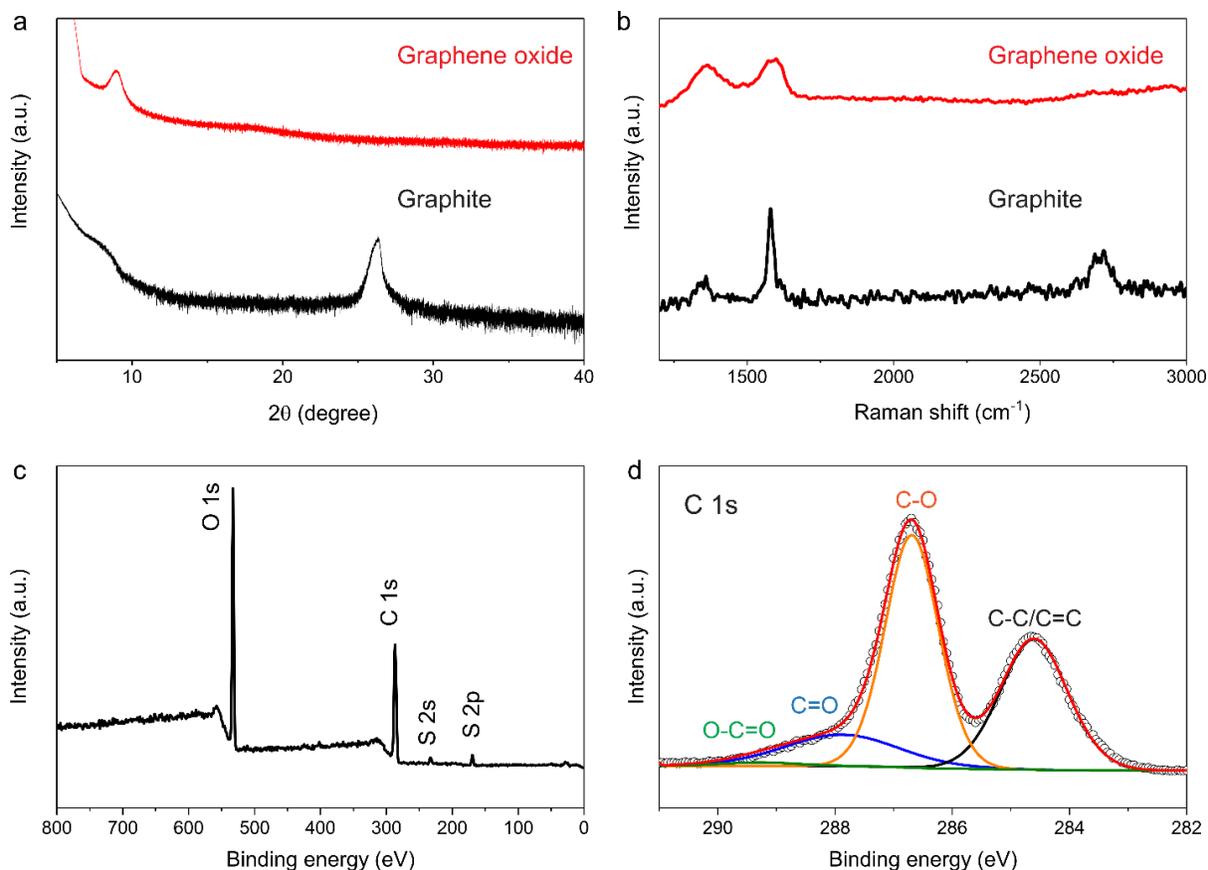

**Fig. S3 Characterization of GO nanosheets.** (a) XRD of graphite and graphene oxide; (b) Raman spectrum of graphite and graphene oxide; (c) XPS spectrum of graphene oxide; (d) High resolution XPS spectrum of C 1$s$.

The XRD pattern of pristine graphite shows a (002) peak with an interlayer spacing of 3.4 Å, while the corresponding (002) peak in graphene oxide shows an interlayer spacing of 9.7 Å. This indicates successful exfoliation of graphite to graphene oxide nanosheets (Fig. S3a). The Raman spectrum of pristine graphite shows a strong and sharp G peak at 1581 cm$^{-1}$ that is attributed to the first-order scattering of the E$_{2g}$ mode[3]. This G peak becomes broader and shifts to 1599 cm$^{-1}$ in graphene oxide. Furthermore, the appearance of a D band at 1365 cm$^{-1}$ indicates the reduction in size of the in-plane $sp^2$ domains (Fig. S3b)[4]. The XPS survey spectra reveal that the main peaks are C and O, the small peaks at a BE of 169.9 eV and 233 eV come from residual S (Fig. S3c). The high-resolution C 1*s* spectrum demonstrates the high oxidation degree of graphene oxide with four components corresponding to C-C/C=C at a BE of 284.8 eV, C-O at 286.9 eV, C=O at 288.1 eV and O-C=O at 289.5 eV (Fig. S3d)[4]. The C 1*s* XPS spectrum of GO confirms a C/O ratio of about 2.1 for GO nanosheets, indicating a high concentration of functional groups that can hydrolyze to generate protons for charging and discharging.

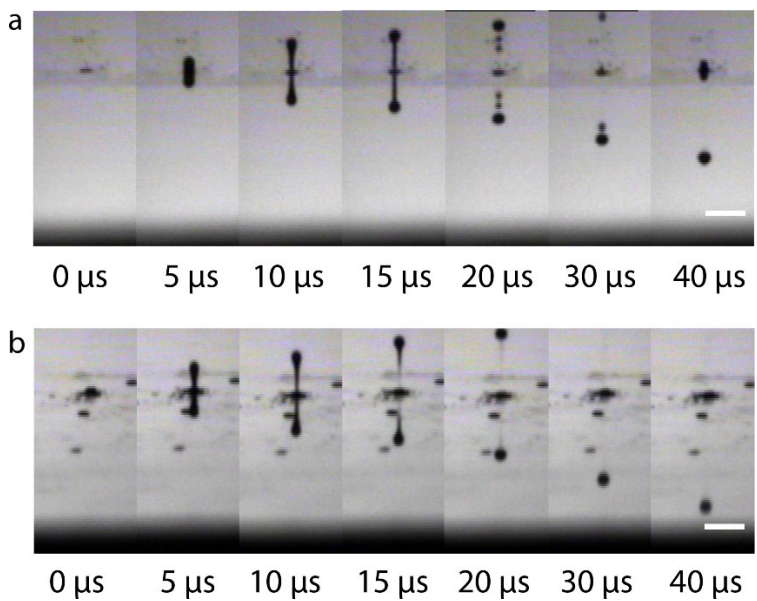

**Fig. S4 Ink stability.** Stroboscopic images of (a) $Ti_3C_2T_x$ and (b) GO ink droplet formation vs. time as observed from the printer camera. The scale bars in (a) and (b) are 50 μm.

To achieve a high quality inkjet printing process, the preparation of printable and stable inks is very important. Water-based MXene ink without any additives showed highly stable printing behavior during jetting, which may be attributed to the presence of functional groups like –O, -OH and -F on the surface of MXene sheets (Fig. S4a) that help dispersion in water. To prepare a printable GO ink, Triton X-100 was added to the water-based graphene oxide ink in order to optimize the ink surface tension[5]. As shown in Fig. S4b, no satellite droplets were generated during jetting, indicating a printable and stable GO ink.

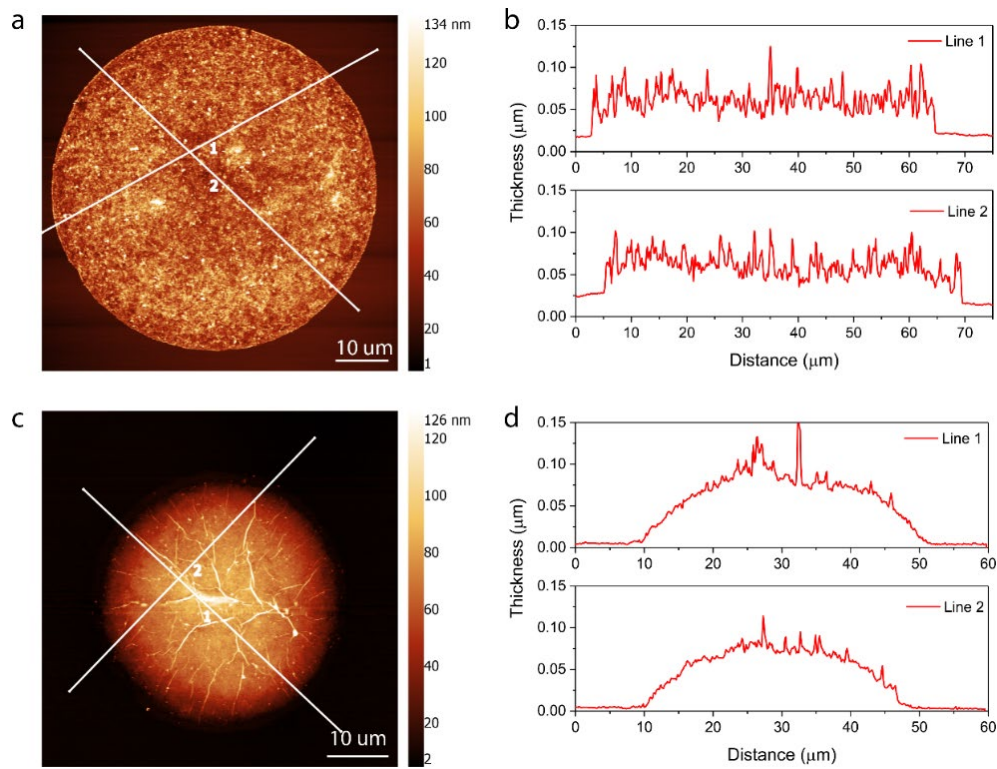

**Fig. S5 AFM analysis of inkjet printed MXene and GO single droplets.** (a) AFM map of single printed $Ti_3C_2T_x$ droplet on silicon substrate; (b) Cross-sectional profiles along two different directions in (a); (c) AFM map of single printed GO droplet on silicon substrate; (d) Cross-sectional profiles along two different directions in (c).

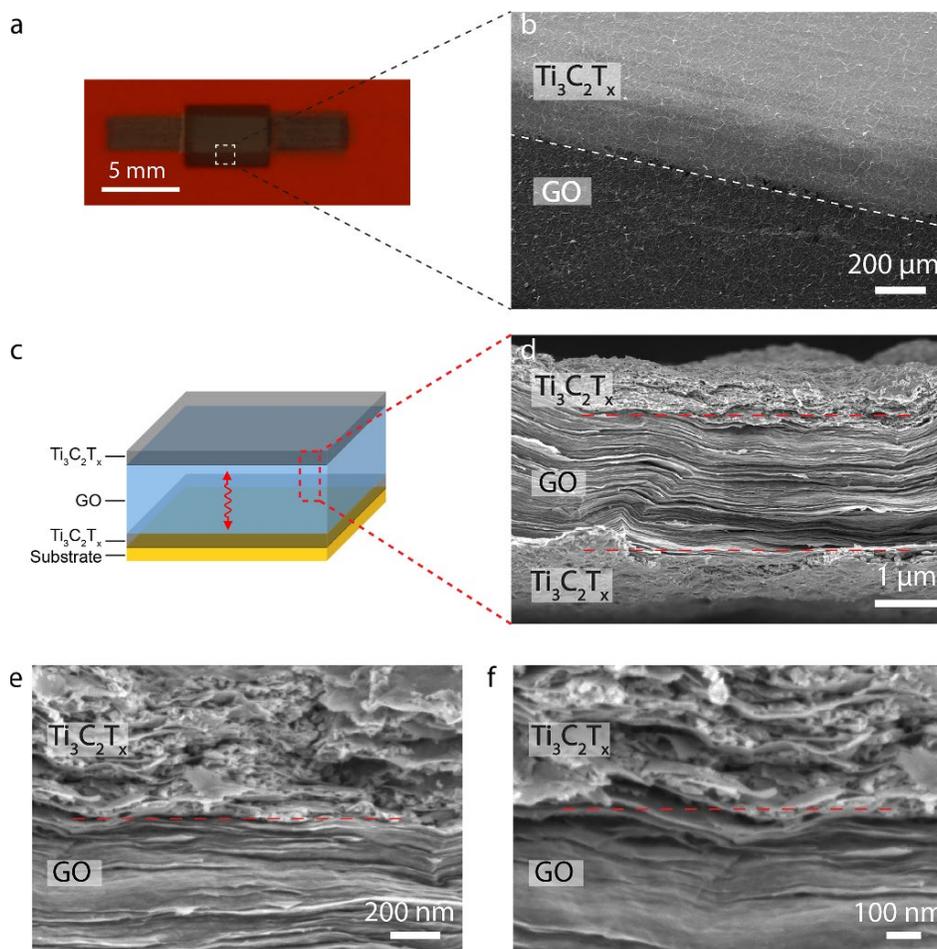

**Fig. S6 30L SSC device morphology.** (a) Optical image of all-inkjet-printed 30L SSCs on flexible polyimide substrate. (b) SEM image of the boundary between $Ti_3C_2T_x$ electrode and GO electrolyte. (c) Schematic representation of sandwiched supercapacitor from networks of nanosheets. (d-f) Cross-sectional SEM images of all-inkjet-printed SSC interfaces. The dashed lines roughly divide $Ti_3C_2T_x$ and GO and serve as a guide to the eye only.

Achieving well controlled and sharp interfaces is a significant challenge for printed heterostructures. The interface plays an important role in device performance. Adding binders such as xanthan gum into inks has been demonstrated to control the structure of the inkjet printed heterostructure interface[5]. However, the performance could also be affected by the presence of

164 (organic) binders. Annealing printed heterostructures at high temperatures can lead to removal of

165 the binder, but it limits the choice of substrates to thermally stable ones. Here, we successfully

166 inkjet printed vertical heterostructures without any sign of re-dispersion at the interface by

167 drying the printed patterns before printing the next layer with different nanosheets. Drying was

168 performed at 50 °C, which is a relatively low temperature and is applicable to most substrates

169 including paper and polymer substrates (Fig. S6a,b). The cross-sectional SEM images a sharp

170 interface with good contact between two different nanosheets (Fig. S6c-f).

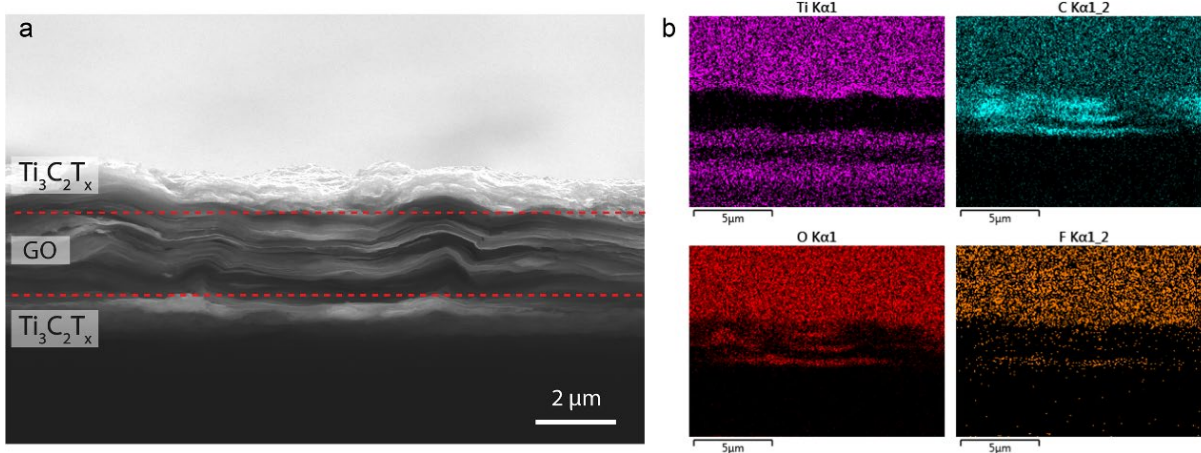

171

172 **Fig. S7 (a) Cross-sectional SEM of all-inkjet-printed SSC and (b) EDS elemental mapping**

173 **of the elements Ti, C, O and F.**

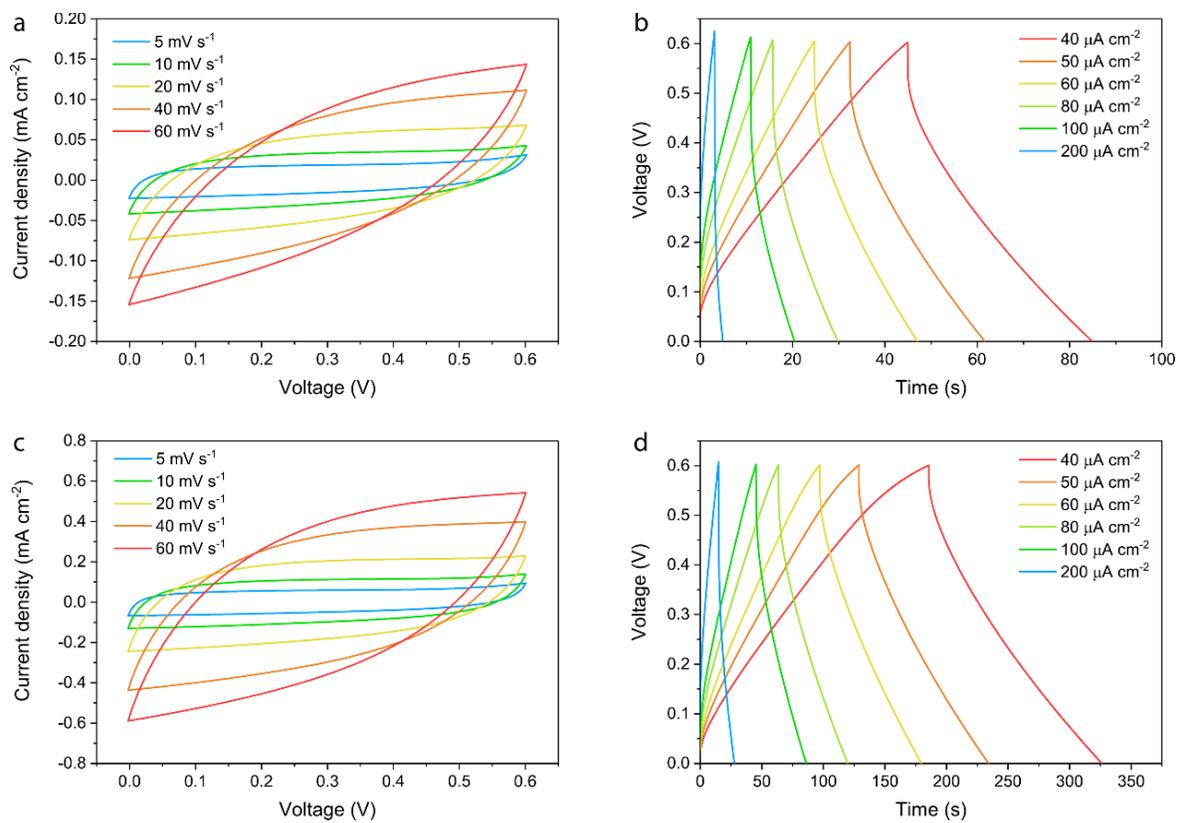

**Fig. S8** CV curves of (a) 10L SSC and (c) 30L SSC as-made devices. GCD curves of (b) 10L SSC and (d) 30L SSC as-made devices at different current densities.

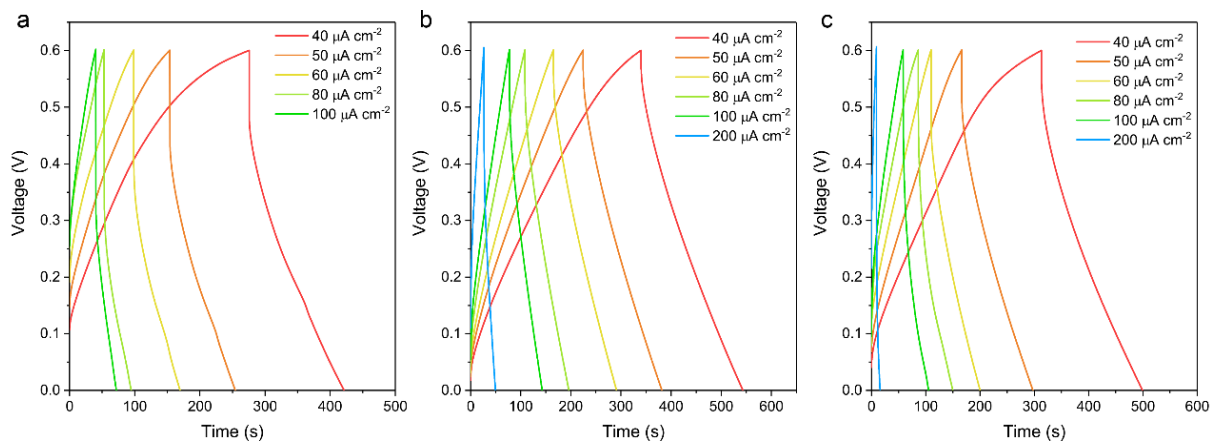

**Fig. S9** GCD curves of 30L SSCs with (a) excess $H_2O$, (b) external aqueous 5 M LiCl electrolyte and (c) aqueous 0.5 M $Na_2SO_4$ electrolyte on top at different current densities.

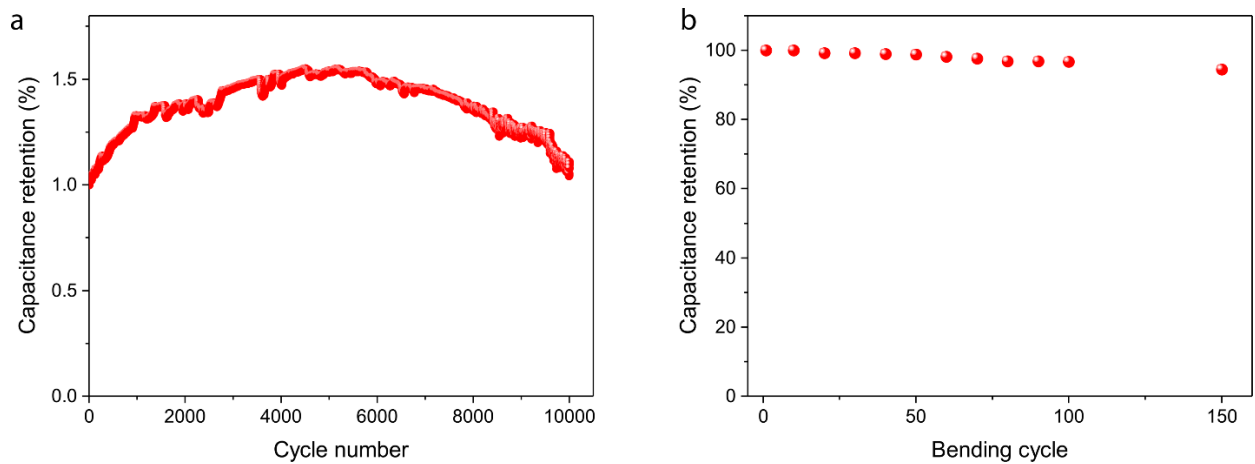

Fig. S10 (a) Cycling stability of 30L SSC. (b) Cyclic bending test of 30L SSC with a bending radius of about 1 cm at a current density of 200 μA cm$^{-2}$ and 94.5% capacitance retention after 150 bending cycles.

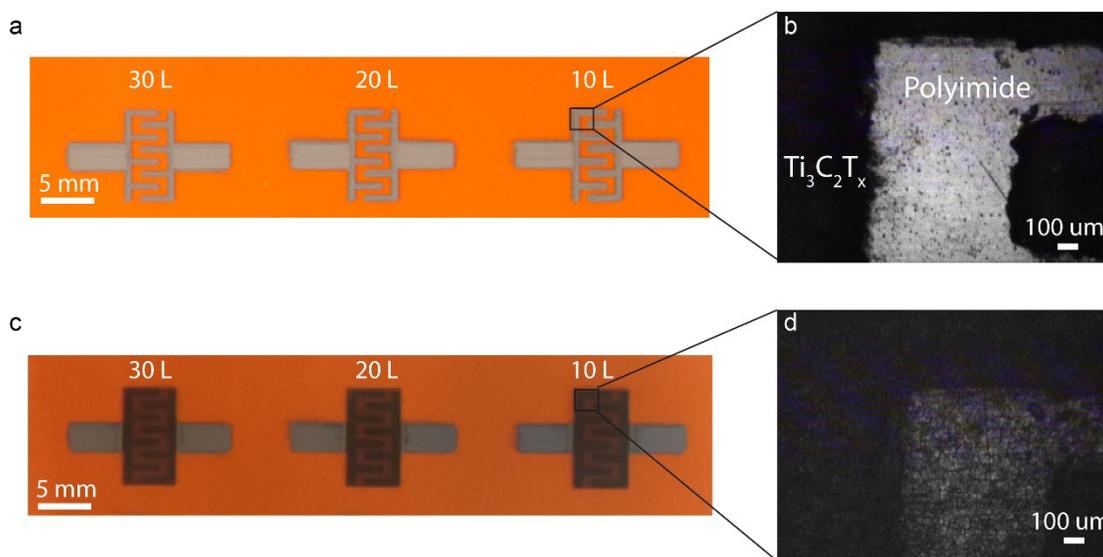

**Fig. S11 Optical image of all-inkjet-printed MSCs.** (a) Optical photograph of printed MXene interdigitated electrodes with different layers on polyimide substrate. (b) Zoom-in photograph of printed electrodes from (a) showing the shape edge between phases. (c) Photograph of complete MSCs with GO electrolyte printed on same interdigitated electrode structure as in (a). (d) Zoom-in image of 30 layer MSCs in (c), indicating full coverage of GO electrolyte on top of MXene interdigitated electrodes. The inset codes in (a) and (c) represent the number of printed layers with which the MXene interdigitated electrodes were formed.

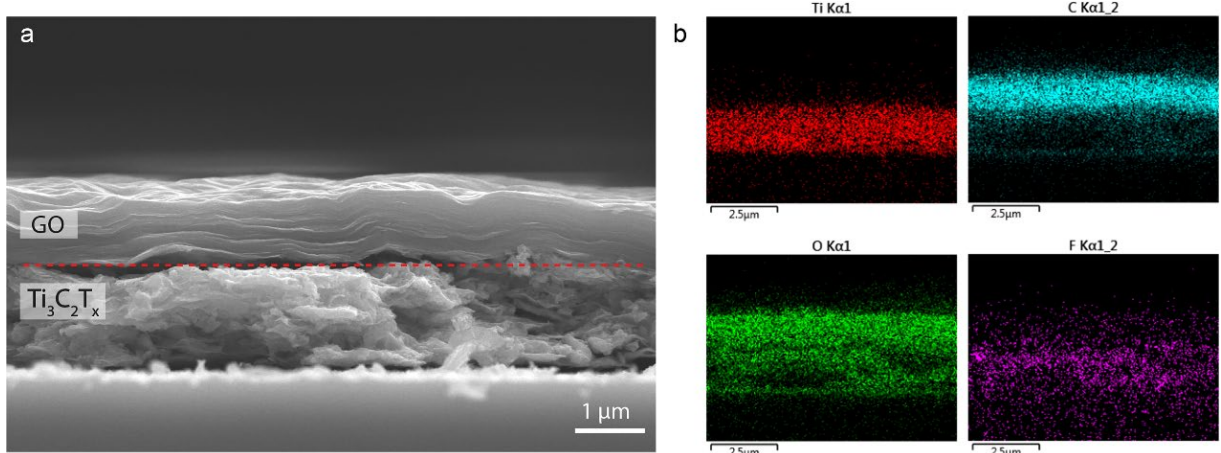

Fig. S12 (a) Cross-sectional SEM image of all-inkjet-printed MSC showing electrode and electrolyte interface and (b) EDS elemental mappings of the elements Ti, C, O and F.

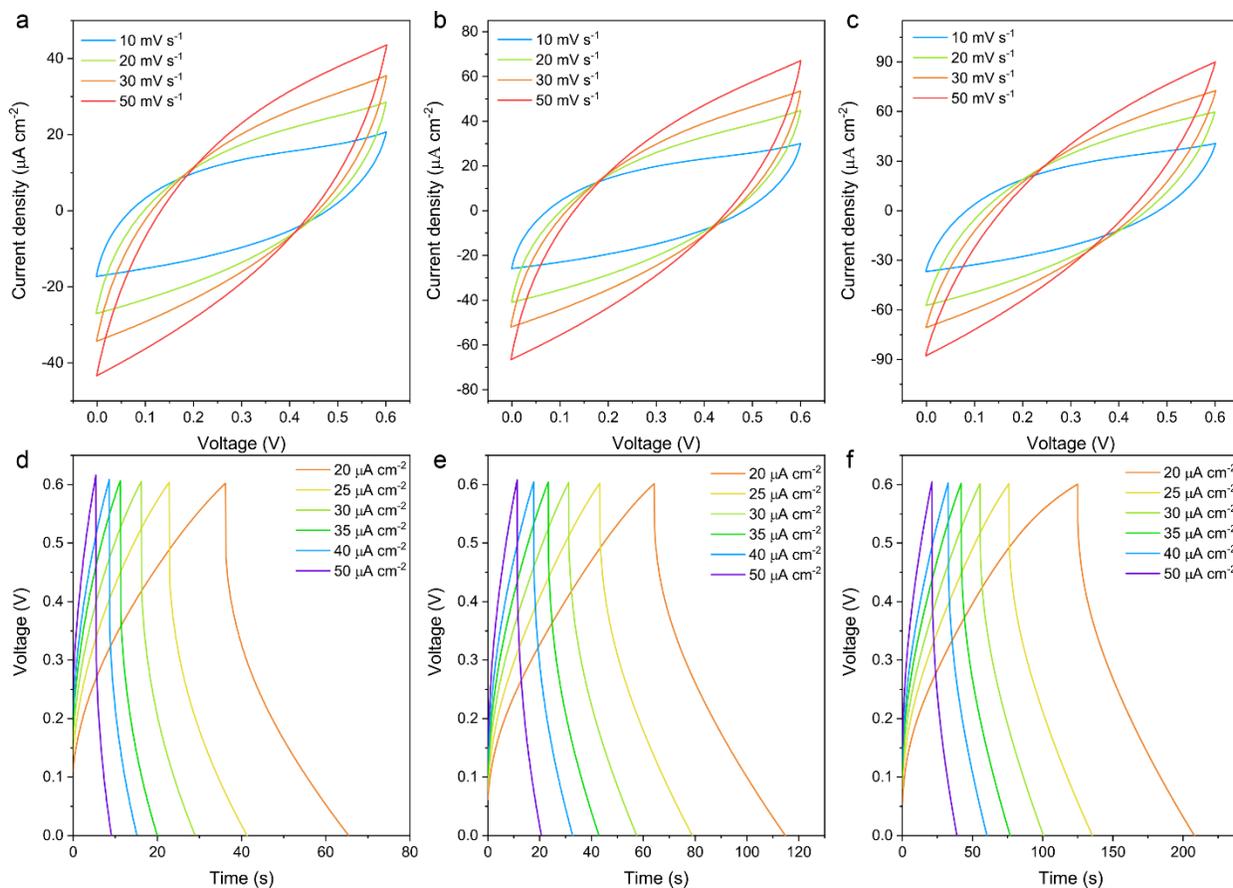

Fig. S13 CV curves of (a) 10L, (b) 20L and (c) 30L MSCs with excess water at different scan rates. GCD curves of (d) 10L, (e) 20L and (f) 30L MSCs with excess water at different current densities.

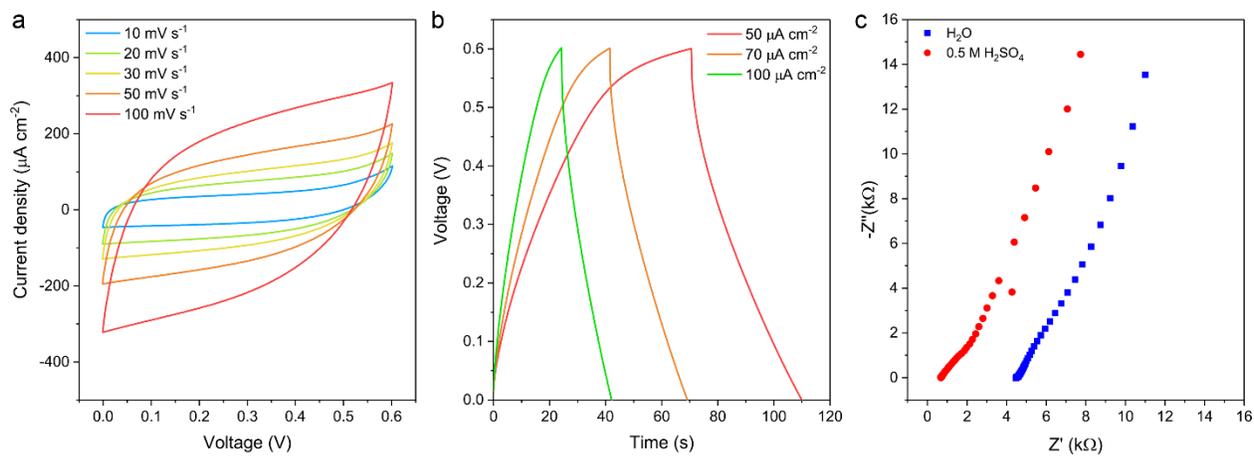

Fig. S14 (a) CV curves of 30L MSC with 0.5 M $H_2SO_4$ electrolyte at different scan rates. (b) GCD curves of 30L MSC with 0.5 M $H_2SO_4$ electrolyte at different current densities. (c) Nyquist plots of 30L MSC with excess water and with 0.5 M $H_2SO_4$ electrolyte.

203 **Table S1. Components of the equivalent circuits fitted for all-inkjet-printed 10L and 30L**
204 **SSCs as shown in Fig. 3f.**

| Samples | $R_1$ (kΩ) | $R_2$ (kΩ) | $Q_1$ (S*s$^n$) ($10^{-4}$) | n | $Q_2$ (S*s$^n$) ($10^{-4}$) | N | $R_3$ (kΩ) | $Q_3$ (S*s$^n$) ($10^{-4}$) | n |
|---|---|---|---|---|---|---|---|---|---|
| 10L SSC | 5.9 | 3.9 | 0.8 | 0.57 | 1.6 | 0.89 | 12.9 | 0.9 | 0.82 |
| 30L SSC | 2.1 | 3.1 | 1.3 | 0.50 | 2.6 | 1.00 | 1.9 | 2.9 | 0.66 |

205

Table S2. Components of the equivalent circuits fitted for all-inkjet-printed MSCs as shown in Fig. 4h.

| Samples | Chi-Squared ($10^{-6}$) | $Q_1$ (S*s$^n$) ($10^{-4}$) | n | $R_1$ (kΩ) | $Q_3$ (S*s$^n$) ($10^{-4}$) | n |
|---|---|---|---|---|---|---|
| 10L MSC | 3.1 | 2.9 | 1.00 | 12.0 | 1.5 | 0.66 |
| 20L MSC | 1.7 | 2.2 | 0.84 | 11.9 | 2.1 | 0.73 |
| 30L MSC | 1.3 | 6.1 | 1.00 | 4.1 | 4.5 | 0.82 |

Due to the complexity in the high frequency range, we only present the low frequency fitting results as shown in Table S2 and Chi-squared which represent the fitting error for high frequency range.